# Gravitational Lensing by Galaxy Quantization States

Franklin Potter [1,2] & Howard G. Preston [3]

**Abstract**    We show how our theory of large-scale gravitational quantization explains the large angle gravitational lensing by galaxies without requiring "dark matter". A galaxy is treated as a collective system of billions of stars in each quantization state with each star experiencing an average gravitational environment analogous to that for nucleons in the atomic nucleus. Consequently, each star is in an approximate finite depth square well type of gravitational potential. The "effective potential" is shown to be about ten times greater than the Newtonian gravitational potential, so the gravitational lensing effects of a galaxy are about ten times greater also, in agreement with the measured gravitational lensing.

[1] Sciencegems.com, 8642 Marvale Drive, Huntington Beach, CA 92646 USA
[2] Please direct correspondence to: drpotter@lycos.com
[3] 15 Vista del Sol, Laguna Beach, CA 92651 USA

## 1. Introduction

In a previous article [1] we introduced a theory of large-scale gravitational quantization and discussed its successful application to the Solar System, to the system of satellites around each of the Jovian planets, to binary systems such as the Earth and Moon, Pluto and Charon, binary stars, and binary galaxies, to galaxies, thereby eliminating the need for "dark matter", and to the accelerated expansion of the universe, with an estimate of a reasonable value for the "dark energy" density. Only two physical parameters of a gravitationally-bound system determine all its quantization states, the total mass and the total vector angular momentum. Yet in spite of such simplicity, we could not find a system which could be considered as the definitive test of the theory because each system investigated has either too large of an uncertainty in the total angular momentum or the Schwarzschild metric is only a rough approximation at best. Therefore, we suggested two possible tests for the future: (1) a laboratory experiment with a rotating massive sphere near to a torsion balance, assuming that the two objects would be gravitationally bound in the horizontal plane, and (2) the measurement of the drift of an Earth satellite toward its predicted equilibrium radius of orbit.

In this article we expand our galaxy application to consider how large-scale gravitational quantization in a galaxy can explain the very large angle gravitational lensing [2] measurements that seem to require a galaxy mass of about ten times its baryonic mass. These lensing measurements have been used as a strong argument for the existence of "dark matter" in galaxies, so the explanation of the gravitational lensing by large-scale quantization is essential for its viability and for eliminating "dark matter".

The theory of large-scale gravitational quantization was derived from the general relativistic Hamilton-Jacobi equation [3] for which quantization conditions are created in a traditional mathematical way. The theory dictates a quantization for each gravitationally-bound system that depends upon two physical quantities of the system only, its total mass M and its total vector angular momentum $H_\Sigma$, each gravitationally-bound system requiring a different system scaling constant

$$H = \frac{H_\Sigma}{M c}, \qquad (1.1)$$

with the quantity $\mu c H$ assuming the traditional role of $\hbar$, where c is the speed of light in vacuum and $\mu$ is the orbiting test particle mass. We showed that large-scale quantization occurs as energy *per mass* and angular momentum *per mass* in terms of H, as derived from the new scalar 'gravitational wave equation' (GWE)

$$g^{\alpha\beta} \frac{\partial^2 \Psi}{\partial x^\alpha \, \partial x^\beta} + \frac{\Psi}{H^2} = 0 \qquad (1.2)$$

The GWE is separable in the Schwarzschild metric (see the Appendix) and the appropriate substitution is the product wave function $\Psi = \Psi_t \Psi_r \Psi_\theta \Psi_\phi$. The radial wave equation becomes

$$\frac{d^2 \Psi_r}{dr^2} + \frac{2}{r} \frac{d\Psi_r}{dr} + \frac{2}{H^2 c^2} \left( \frac{E}{\mu} + \frac{r_g c^2}{2r} - \frac{\ell(\ell+1) H^2 c^2}{2r^2} \right) \Psi_r \approx 0, \qquad (1.3)$$

which is Schrödinger-like and similar to the radial wave equation for the hydrogen atom. The Schwarzschild radius $r_g = 2GM/c^2$. Instead of the electrostatic potential, this equation has the gravitational potential as the middle term in the large bracket, and $\mu c H$ replaces $\hbar$. The energy for a quantization state with principal quantum number $n$ is

$$E_n = -\mu c^2 \frac{r_g^2}{8 n^2 H^2}, \qquad (1.4)$$

on the order of $10^{-6} \, \mu c^2$ or smaller for most cases of interest. We see that the energy *per mass* is quantized. We see also that when $n = 1$ for the system in the Schwarzschild metric consisting of a central mass and an orbiting body, there is a minimum energy state, unlike the classical case.

For simplicity, we concentrate on circular or near-circular orbits only. When $\ell = n - 1$, the orbit is circular with a single radial peak in the probability distribution. We define a "gravitational Bohr radius" $r_0 = 2H^2/r_g$, which establishes the distance scale for the gravitationally-bound system because the peak in the wave function probability occurs at $n^2 r_0$ for each $n$. There is also an equilibrium radius of orbit $r_{eq} = n(n-1) r_0$ calculated from the negative gradient of the kinetic energy bracket in the radial wave equation. And finally, one can define a 'gravitational de Broglie' wavelength $\lambda = 2\pi n \, 2H^2/r_g$, which is independent of the mass of the particle in orbit in agreement with the equivalence principle.

We showed also that there is an uncertainty principle in this large-scale quantization theory for the *gravitational* interaction

$$\Delta x \, \Delta p \geq \mu c H / 2. \qquad (1.5)$$

and that a classical description of the continuous particle trajectory is allowed in the theory as long as gravitons are not required for the observation.

## 2. Gravitational Lensing by Galaxies

Previously, we applied the large-scale quantization theory to the Galaxy by utilizing the GWE solutions for the Schwarzschild metric as a rough approximation because the actual metric for the disk does not lead to a separable differential equation. That is, we considered the quantization state of a test particle (i.e., a star) in orbit *outside* the visible extent of the galactic disk at a distance slightly greater than its luminous baryonic mass radius R ~ 5 x $10^{20}$ m and assumed that the Schwarzschild metric solutions of the GWE are a reasonable approximation for this situation. Then the energy eigenvalue is given still by equation (1.4). The application of the virial theorem led directly to a tangential rotation velocity for the test particle

$$v = \frac{r_g c}{2 n H}. \qquad (2.6)$$

We found that by using the reasonably low values of M = 6 x $10^{10}$ solar masses and $H_\Sigma$ ~ 4.4 x $10^{66}$ kg-m$^2$ s$^{-1}$ we could accommodate the measured [4] rotational velocity v ~ 2.2 x $10^5$ m s$^{-1}$ with $n = 1$. The 'dark matter' model requires a total Galaxy mass of at least 7.2 x $10^{11}$ solar masses, with about 10% being attributed to luminous baryonic matter. For the $n = 2$ eigenstate, the theory predicts a velocity that is one-half as much or 1.1 x $10^5$ m s$^{-1}$, in agreement with the recent values [5] for the velocities of the streaming stars just beyond the edge of the visible disk. From these results, we stated that the Galaxy is not a classical system but, instead, must be described by a large-scale quantization theory.

We continued to discuss the application of the theory to a galaxy, deriving the baryonic Tully-Fisher relation and the MOND acceleration [2, 6] in terms of M and $H_\Sigma$. And we stated: "If we were to develop a more detailed model of the Galaxy, we would abandon the Schwarzschild metric approximation and would approximate the galactic potential in the disk region by a nuclear shell potential of nuclear physics that averages the contribution from all the particles. One can show that at most two or three bound states survive in the shell-model potential well, the states with quantum numbers $n = 1$, $\ell = 0$ and $\ell = 1$ and possibly $n = 2$, $\ell = 0$. Then one would interpret the $n = 1$, $\ell = 0$ state as corresponding to the galactic nucleus because having $\ell = 0$ would allow mass to accumulate at small radius into a sphere. The $n = 1$, $l = 1$ state would correspond to the disk state with matching to the Schwarzschild state at its edge. The $n = 2$ state would contain the streaming stars beyond the disk. The same approach would be applied to a cluster of galaxies to determine its quantization states. We leave these developments for future research."

In contrast to the Solar System type of gravitationally bound systems in which one or perhaps a few bodies occupy each quantization state, galaxies have billions of bodies in each quantization state. Each massive body experiences the average collective gravitational potential of all the other bodies, much like the environment each nucleon experiences in an atomic nucleus. So let's try a nuclear shell model type of potential well and compare its predictions with the physical properties of a galaxy, particularly its gravitational lensing behavior. In the radial equation (1.3) we can substitute a general potential V(r) in place of the Newtonian potential to obtain

$$\frac{d^2 \Psi_r}{dr^2} + \frac{2}{r}\frac{d\Psi_r}{dr} + \frac{2}{H^2 c^2}\left(\frac{E}{\mu} - \frac{V(r)}{\mu} - \frac{\ell(\ell+1) H^2 c^2}{2 r^2}\right)\Psi_r \approx 0. \tag{2.7}$$

For galaxies, we now assume a finite depth square well potential of "effective radius" b and constant depth $V(r) = -V_0$. The confluent hypergeometric functions of our earlier solutions now become spherical Bessel functions

$$\Psi_r^{n\ell} \sim \frac{A_\ell \, J_{\ell+1/2}[k_{n\ell}\, r]}{\sqrt{r}}, \tag{2.8}$$

where $k_{n0} = (\sqrt{2(E+V_0)/\mu})/cH$ for $\ell = 0$, with $E < 0$. A second solution with the Bessel Y function also may apply. For $n = 1$, we can have $\ell = 0, 1, 2$, etc., but the quantization states with $\ell > 2$ will have a higher energy than the $n = 2$, $\ell = 0$ state. The quantity $k_{n0} = n\pi/b$, therefore each $\ell = 0$ quantization state has

$$\frac{\sqrt{2(E+V_0)/\mu}}{cH} = n\pi/b, \tag{2.9}$$

i.e., an energy

$$E_{n0} = \mu c^2 \left(\frac{n^2 \pi^2 H^2}{2 b^2} - \frac{V_0}{\mu c^2}\right) \tag{2.10}$$

which leads to a bound state when $V_0$ is deep enough. Equating this quantization state energy for $n = 1$ to the energy for the Schwarzschild approximation given in equation (1.4) produces

$$V_0 = \mu c^2 \left(\frac{r_g^2}{8 H^2} + \frac{\pi^2 H^2}{2 b^2}\right), \tag{2.11}$$

the first term being equivalent to the Newtonian gravitational potential at the "gravitational Bohr radius" $r_0 = 2H^2/r_g$. We can set $b = \beta r_0$, with the result being

$$V_0 = \mu c^2 \frac{r_g^2}{8 H^2}\left(1 + \frac{\pi^2}{\beta^2}\right). \tag{2.12}$$

If $\beta \sim 1$, then the "effective potential" of the large-scale gravitational quantization theory is about ten times the standard Newtonian potential. For the Galaxy (Milky Way), using the values listed above for M and $H_\Sigma$, $r_0 \sim 1.7 \times 10^{20}$ m, placing the approximate radius b of the potential well within the disk. If we use the larger mass of 1.6 x $10^{11}$ solar masses, a traditional Newtonian value, with its corresponding $H_\Sigma \sim 3.3 \times 10^{67}$ kg-m$^2$ s$^{-1}$, then b $\sim 5 \times 10^{20}$ m, a distance just beyond the edge of the visible disk, probably a more reasonable potential well radius. The point here is that one can accommodate $\beta \sim 1$ with reasonable M and $H_\Sigma$ values for the expected well radius without including any "dark matter".

The angular deflection $\theta$ of a ray of light from its undeviated path produced by gravitational lensing depends upon the gravitational potential in the approximate form

$$\theta \sim 4 \frac{GM}{r c^2}, \tag{2.13}$$

so if the effective potential is about ten times greater than the Newtonian potential GM/r at each value of r, then the angle of deflection will be about ten times greater. The integral over the gradient of the potential for a mass distribution within the light ray approach distance r will yield the same conclusion.

Therefore our large-scale gravitational quantization theory predicts the large gravitational lensing angular deflection values in reasonable agreement with the measured values. There is no need to invoke "dark matter" as the cause of these large deflections of light or for the nearly constant rotational velocity of the disk stars. We conclude again that galaxies behave as large-scale quantization systems.

## APPENDIX: Review of Large-scale Gravitational Quantization Theory

The general relativistic Hamilton-Jacobi equation for a test particle of mass $\mu$ as given by Landau and Lifshitz [3] is

$$g^{\alpha\beta} \frac{\partial S}{\partial x^\alpha} \frac{\partial S}{\partial x^\beta} - \mu^2 c^2 = 0, \tag{A1}$$

where $g^{\alpha\beta}$ is the metric of the general theory of relativity and S is the action. This relativistic Hamilton-Jacobi equation then becomes a wave equation via the transformation to eliminate the squared first derivative, i.e., by defining the wave function $\Psi$ (q, p, t) of position q, momentum p, and time t

$$\Psi = e^{iS'/H}. \tag{A2}$$

with S' = S/$\mu$c. The end result is the new scalar 'gravitational wave equation' (GWE)

$$g^{\alpha\beta} \frac{\partial^2 \Psi}{\partial x^\alpha \partial x^\beta} + \frac{\Psi}{H^2} = 0 \tag{A3}$$

The GWE is separable in the Schwarzschild metric and the appropriate substitution is the product wave function $\Psi = \Psi_t \Psi_r \Psi_\theta \Psi_\phi$. Recall that $S = -E_0 t + S(r) + S(\theta) + L\phi$. Separation of variables produces the coordinate equations, with the primes representing division by $\mu c$:

$$\frac{d^2 \Psi_t}{c^2 \, dt^2} = \frac{-E_0^{'2}}{H^2 c^2} \Psi_t \tag{A4}$$

$$\frac{d^2 \Psi_\phi}{d\phi^2} = \frac{-L^{'2}}{H^2} \Psi_\phi \tag{A5}$$

$$\frac{1}{\sin\theta} \frac{d}{d\theta}\left(\sin\theta \frac{d\Psi_\theta}{d\theta}\right) - \frac{m^2}{\sin^2\theta} \Psi_\theta = -\ell(\ell+1) \Psi_\theta \tag{A6}$$

$$\left(1 - \frac{r_g}{r}\right) \frac{d^2 \Psi_r}{dr^2} + \frac{(2 - \frac{r_g}{r})}{r} \frac{d\Psi_r}{dr} + \left(\left(1 - \frac{r_g}{r}\right)^{-1} \frac{E_0^{'2}}{H^2 c^2} - \frac{1}{H^2} - \frac{\ell(\ell+1)}{r^2}\right) \Psi_r = 0 \tag{A7}$$

The relativistic energy $E_0'$ and the angular momentum $L'$ are associated with the particle in orbit in the gravitationally-bound system.

Requiring $\Psi_\phi$ to be a single-valued function dictates the angular momentum quantization condition

$$L'/H = m \tag{A8}$$

with integer $m = 0, \pm 1, \pm 2$, etc. Because $L' = L/\mu c$, the angular momentum *per mass* is quantized. The equation for $\Psi_\theta$ determines the azimuthal quantum number $\ell = 0, 1, 2$, etc., with $|m| \le \ell$. When $|m| = \ell$ the maximum probability occurs at $\theta = \pi/2$, i.e., around the equatorial plane, where most of the orbiting mass for gravitationally-bound systems lies.

This radial equation has characteristics of a Klein-Gordon equation, so one cannot guarantee in general that the wave function $\Psi$ always has positive probability. For the electromagnetic interaction with its positive and negative electrical charges and currents, this failing of the Klein-Gordon equation is remedied by using the charge density and the current density instead of the probability density. For the gravitational interaction, even though mass and energy are equivalent, we are not sure that a negative mass density and negative mass current density are appropriate physical quantities. Therefore, we progress immediately to a Schrödinger-like equation that is first order in energy.

The radial equation has a singularity at $r = r_g$ that is transformed away by the standard substitution $r(r - r_g) = r'^2$, where $r'$ is a new radial coordinate. In all cases of interest, $r_g \ll r'$, typically $r_g/r' < 10^{-8}$, so we choose to ignore terms proportional to $r_g/r'^2$, $r_g^2/r'^2$, and smaller. We make the traditional substitution for the relativistic energy $E_0 = \mu c^2 + E$, with $E \ll \mu c^2$, so the radial equation becomes

$$\frac{d^2 \Psi_{r'}}{dr'^2} + \frac{2}{r'} \frac{d\Psi_{r'}}{dr'} + \frac{2}{H^2 c^2} \left(\frac{E}{\mu} + \frac{r_g c^2}{2 r'} - \frac{\ell(\ell+1) H^2 c^2}{2 r'^2}\right) \Psi_{r'} \approx 0, \tag{A9}$$

which is Schrödinger-like and similar to the radial wave equation for the hydrogen atom. Instead of the electrostatic potential, this equation has the gravitational potential as the middle term in the large bracket, and $\mu c H$ replaces $\hbar$. The solution for $\Psi_r$ is, after dropping the prime on the r for simplicity,

$$\Psi_r = A r^\ell \exp\left(\frac{-\sqrt{-2E/\mu}}{Hc} r\right) {}_1F_1\left(\frac{-r_g c}{2H\sqrt{-2E/\mu}} + \ell + 1, 2\ell + 2; 2\frac{\sqrt{-2E/\mu}}{Hc} r\right) \tag{A10}$$

with A the normalization constant. The quantity $\Psi\Psi^\dagger$ gives the probability density of the particle.

If the first parameter in the confluent hypergeometric function ${}_1F_1(g, h; z)$ is a negative integer or zero, then ${}_1F_1(g, h; z)$ reduces to a polynomial with a finite number of terms and the wave function does not diverge at infinity. Additionally, its second parameter must be a positive integer for the function to be single-valued. The second solution with $U(g, h; z)$ does not yield a satisfactory wave function [7].

Setting the first parameter in $_1F_1$ (g, h; z) equal to $-n_r$, a negative integer or zero, and solving for the energy produces

$$E_n = -\mu c^2 \frac{r_g^2}{8 n^2 H^2},\qquad(A11)$$

on the order of $10^{-6}\ \mu c^2$ or smaller for most cases of interest. The principal quantum number $n = n_r + \ell + 1$, always a positive integer.